\documentclass[aip, rsi, amsmath, amssymb, reprint, author-year, floatfix]{revtex4-1}
\usepackage{amsmath}
\usepackage{graphicx}
\usepackage[caption=false]{subfig}
\usepackage{lineno}
\usepackage{color}
\usepackage{hyperref}

\begin{document}
	
\preprint{AIP/123-QED}
	
\title{Calorimetric observation of single He$_2^*$ excimers in a 100~mK He bath}
\author{F.W. Carter}
\email{faustin.carter@gmail.com}
\affiliation{Argonne National Laboratory, High Energy Physics, Lemont, IL, 60439}
\affiliation{Yale University, Department of Physics, New Haven, CT 06511}

\author{S.A. Hertel}
\affiliation{University of California Berkeley, Department of Physics, Berkeley, CA 94720}
\affiliation{Lawrence Berkeley National Laboratory, Berkeley, CA 94720}
\affiliation{Yale University, Department of Physics, New Haven, CT 06511}

\author{M.J. Rooks}
\affiliation{Yale Institute for Nanoscience and Quantum Engineering, New Haven, CT 06520}

\author{P.V.E. McClintock}
\affiliation{Department of Physics, Lancaster University, Lancaster LA1 4YB, United Kingdom}

\author{D.N. McKinsey}
\affiliation{University of California Berkeley, Department of Physics, Berkeley, CA 94720}
\affiliation{Lawrence Berkeley National Laboratory, Berkeley, CA 94720}
\affiliation{Yale University, Department of Physics, New Haven, CT 06511}

\author{D.E. Prober}
\affiliation{Yale University, Department of Applied Physics, New Haven, CT 06520}

\date{\today}

\begin{abstract}
We report the first calorimetric detection of individual He$_2^*$ excimers within a bath of superfluid $^4$He. The detector used in this work is a single superconducting titanium transition edge sensor (TES) with an energy resolution of $\sim$1~eV, immersed directly in the helium bath. He$_2^*$ excimers are produced in the surrounding bath using an external gamma-ray source. These excimers exist either as short-lived singlet or long-lived triplet states. We demonstrate detection (and discrimination) of both states: in the singlet case the calorimeter records the absorption of a prompt $\approx$15~eV photon, and in the triplet case the calorimeter records a direct interaction of the molecule with the TES surface, which deposits a distinct fraction of the $\approx$15~eV, released upon decay, into the surface. We also briefly discuss the detector fabrication and characterization.
\end{abstract}


\maketitle

\section{Introduction}
Superfluid helium, when subjected to ionizing radiation, produces metastable diatomic He molecules in both the singlet and triplet states, emitting a $\approx$15~eV photon upon decay. The singlet He$_2^*$(A$^1\Sigma_u^+$) decays within nanoseconds, while the triplet He$_2^*$(a$^3\Sigma_u^+$) exhibits a remarkably long lifetime of 13 seconds in superfluid helium \citep{McKinsey1999}. The long-lived triplet state can serve as an observable tracer particle in a liquid helium bath, tagging the flow of the normal-fluid component or, at colder temperatures, tagging quantized vortices in the superfluid component \citep{Guo2014}. Efficient detection of helium excimers may also enable the use of a superfluid helium bath in a search for dark matter-induced nuclear recoils, given that a recoil's resulting singlet:triplet excimer ratio distinguishes between electron- and nuclear-recoils. Additionally, high-sensitivity detection of electronic excitations may be used to veto electron recoil backgrounds when searching for low energy nuclear recoils, since nuclear recoils predominantly produce heat in the form of rotons and phonons \citep{Guo2013}.

These applications require an efficient technique for observing and differentiating between the two excimer states. Two techniques have been employed previously for triplet excimer detection: observation through laser fluorescence~\citep{Rellergert2008, Guo2009}, and observation of Auger electrons produced through quenching on a surface~\citep{Zmeev2012, Zmeev2013}. Low-temperature calorimetry offers near-unity efficiency for detecting \emph{any} energy deposition above some energy threshold, making both the $\approx$15~eV singlet decay photons and the triplet surface quench process observable with the same sensor. Here we report success at this high-efficiency detection of both excimer states by a single sensor. 

In this work, we employ a single Transition Edge Sensor (TES) with resolution of $~\sim 1$~eV at 15~eV, similar to a prototype discussed in earlier work ~\citep{Carter2015}. The detector is immersed directly in a superfluid helium bath at 100~mK (the Kapitza resistance between the sensor and liquid helium allows the sensor to function even while in contact with the superfluid helium), and excimers are created by exposing the bath to gamma-rays from a 100~$\mu$Curie $^{22}$Na source located outside the cryostat. The superfluid reservoir, the cryostat, and the electronics are described in \citet{Carter2015a}.

\section{The Transition Edge Sensor}

\begin{table}
	\renewcommand{\arraystretch}{1.5}
	\caption{Characteristics of the Ti TES with Al leads and an integrated 100~nm thick Al/Cu thin-film aperture.}
	\label{tab:tes-params}
	\centering
	\begin{tabular}{l c | l c}
		\hline
		\textbf{Physical} & ~ & \textbf{External} & ~  \\
		\hline
		Width & 10~$\mu$m & $L_\mathrm{total}$ & 50~nH\\
		Length & 15~$\mu$m &  $R_\mathrm{shunt}$ & 200~m$\Omega$\\
		Thickness & 15~nm & $R_\mathrm{parasitic}$ & 5~m$\Omega$\\
		$R_\mathrm{Normal}$ & 48.6~$\Omega$ & $T_\mathrm{bath}$ & 100~mK  \\
		$T_\mathrm{c}$ & 345~mK & $I_\mathrm{bias}$ & 18~$\mu$A\\
		\hline
	\end{tabular}
\end{table}

\begin{figure*}[!th]
	\subfloat[]{\includegraphics[height=1.5in]{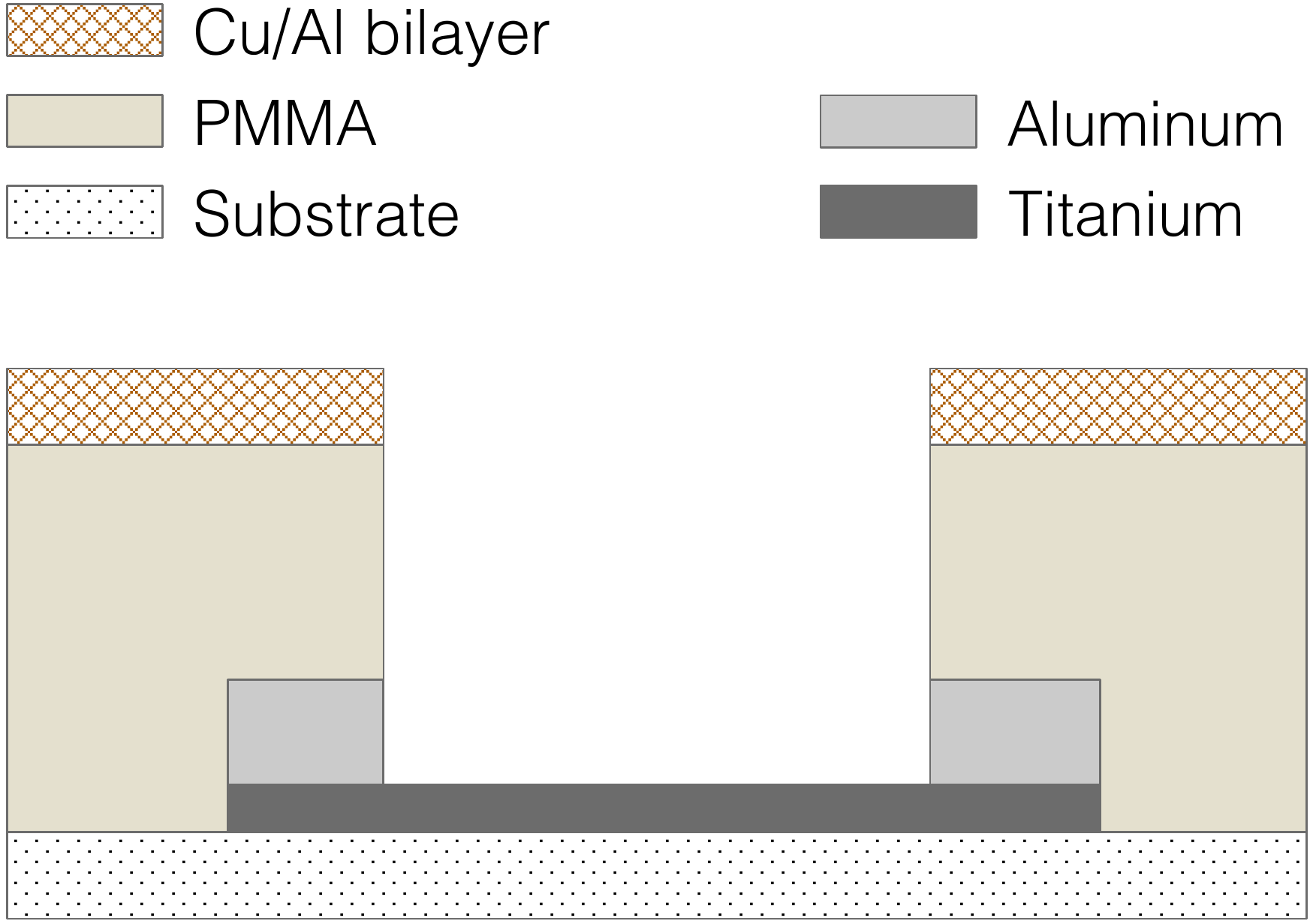}\label{fig:tes-cartoon}}
	\hfil
	\subfloat[]{\includegraphics[height=1.5in]{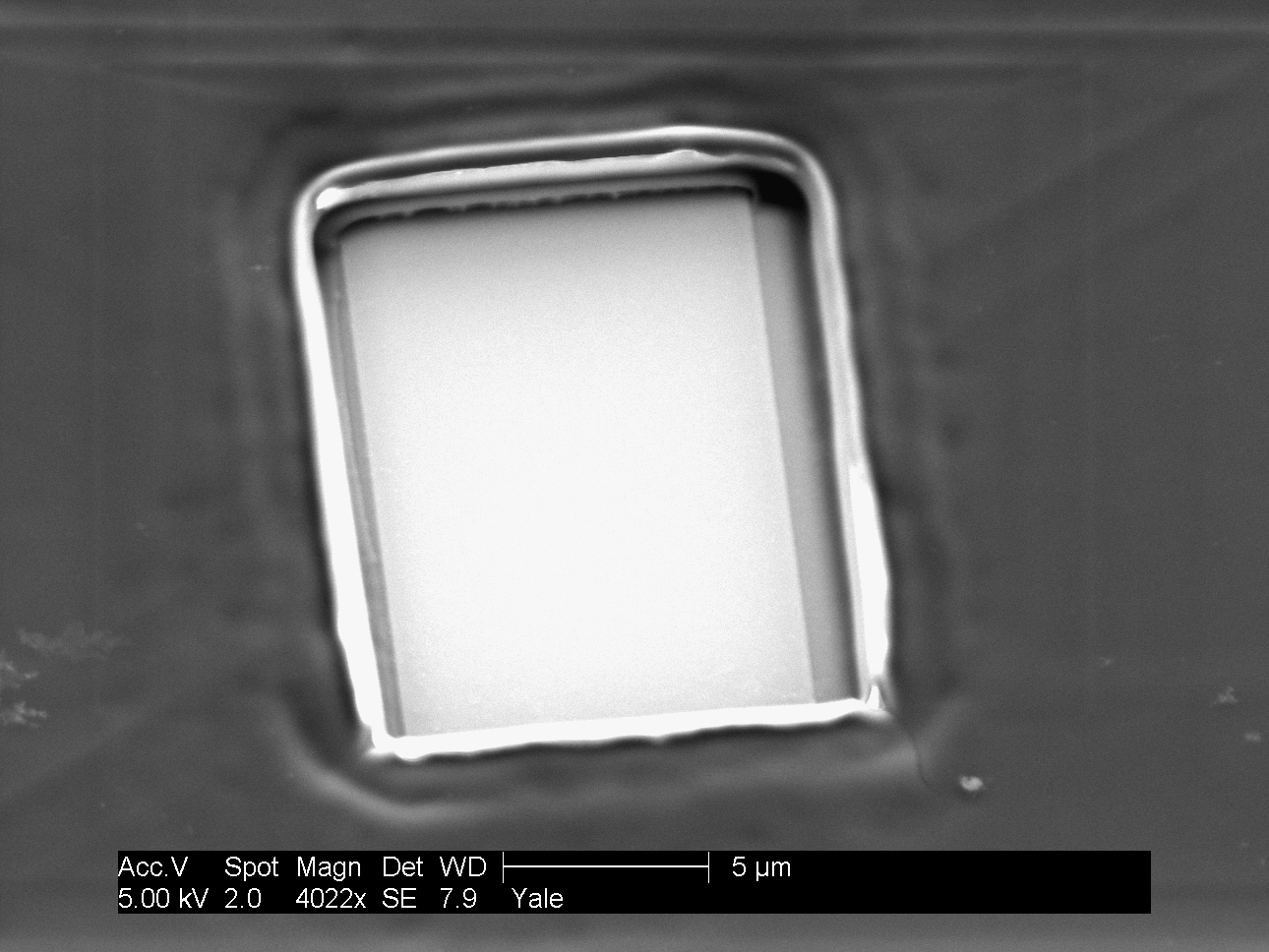}\label{fig:tes-sem}}
	\hfil
	\subfloat[]{\includegraphics[height=1.5in]{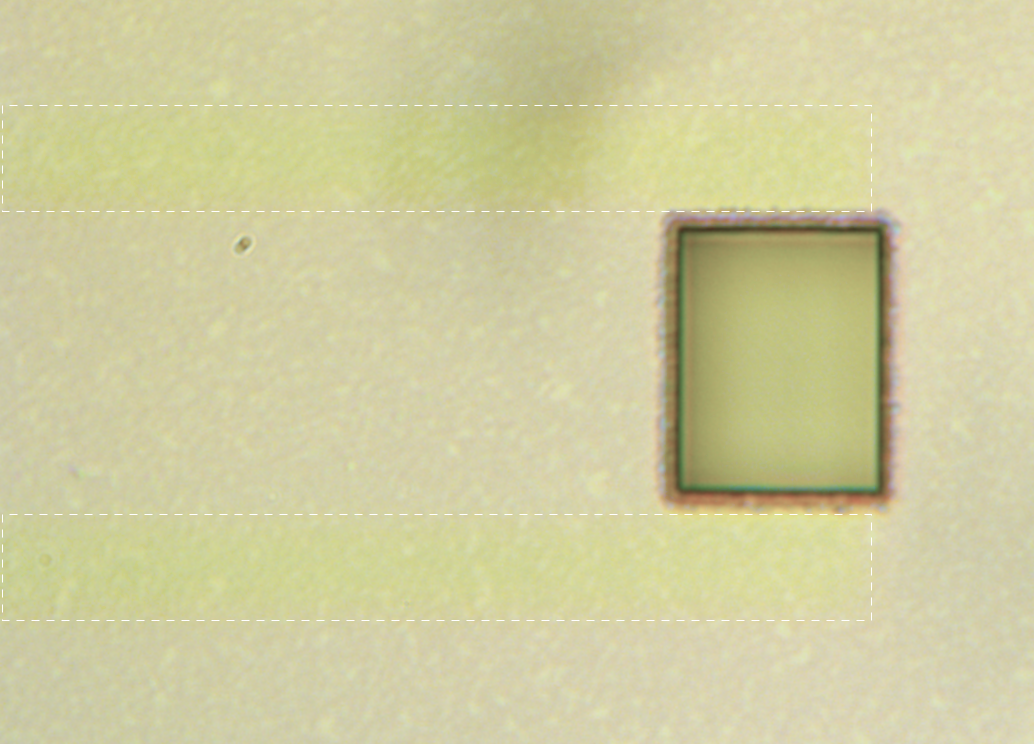}\label{fig:tes-optical}}
	\caption{(a) Cross-sectional diagram of TES configuration (not to scale, dimensions in table~\ref{tab:tes-params}). (b) SEM image of TES with aperture. (c) Optical image of TES with aperture. Leads are just visible extending to the left underneath the Cu/Al shield layer (outlined in dashed white).}
	\label{fig:tes-images}
\end{figure*}

The TES consists of a 15~nm film of evaporated titanium with 300~nm thick evaporated aluminum leads and an integrated 100~nm thick Cu/Al aperture. Table \ref{tab:tes-params} gives basic device parameters. The dominant cooling mechanism, which sets the detector time constant of $\approx$800~ns, is electron-phonon coupling in the Ti (electron outdiffusion is blocked by Andreev reflection in the superconducting Al leads). The TES is operated in negative electro-thermal feedback mode by wiring it in parallel with a small shunt resistor (see table~\ref{tab:tes-params}) and providing a current bias~\citep{Irwin1995}. The current through the TES is read out with a SQUID amplifier and room-temperature electronics from Magnicon~\citep{Drung2007}, which are coupled to a fast (6~GHz) digitizing oscilloscope. Energy deposited in the voltage-biased TES results in a negative current pulse with an integrated charge that is proportional to the energy absorbed. Henceforth, we refer to these pulses as `events'. Event pulses are recorded, filtered, and fit to a model-pulse derived from the TES response to several thousand single blue photon events. The integral of the resulting best-fit pulse is then scaled by the TES efficiency to give our best estimate of the incident energy.

The TES is protected by a thin-film aperture intended to absorb and diffuse energy deposited near, but not directly into the TES. The aperture is fabricated by evaporating a 100~nm thick Cu/Al bi-layer on top of a thick 1~$\mu$m layer of insulating polymethylmethacrylate (PMMA) spun directly onto the wafer, and then etching a window directly over the TES. The full fabrication process of the TES is detailed in the appendix and Fig.~\ref{fig:tes-cartoon} shows a side view of the layers. Figures \ref{fig:tes-optical} and \ref{fig:tes-sem} show optical and SEM images respectively.

We characterized our detector response with a pulsed blue (2.6~eV) laser, in the same manner as in \citet{Carter2015}. By illuminating the TES with a pulsed source of low average photon number and making a histogram of the measured TES pulse-areas, one may calculate the TES energy resolution and efficiency by fitting the histogram with a Gaussian-broadened Poisson distribution.

\section{Experimental setup}
The TES was mounted on the inner wall of a $5\times5\times5$~cm$^3$ helium-filled chamber, held at a temperature of 100~mK. The helium was pure $^4$He (less than 1 part in 10$^{12}$ $^3$He) produced at Lancaster University using the heat-flush technique \citep{McClintock1978}. Excimers were created in the helium bath in two ways. The first method employed a $^{22}$Na gamma-ray source external to the cryostat. Compton scattering in the helium produces electron recoils extending to hundreds of keV. These high-energy electrons lose their energy to the surrounding He atoms by exciting and ionizing He atoms, which upon electron-ion recombination produce a mixture of singlet and triplet excimers. The second method for excimer production was to apply a large negative voltage ($\sim$-1.5~kV) to a sharp tungsten tip immersed in the He bath. Electrons emitted from the tip lose energy to the bath, producing atomic excitations, rotons, and quantized vortex rings. Both production methods were employed with consistent results. Here we report results from the $^{22}$Na method, which allowed a cleaner timing selection (through the use of a coincidence trigger as explained below) and lower electronic readout noise.

\begin{figure}[h]
	\subfloat[No helium]{\includegraphics[width=3.25in]{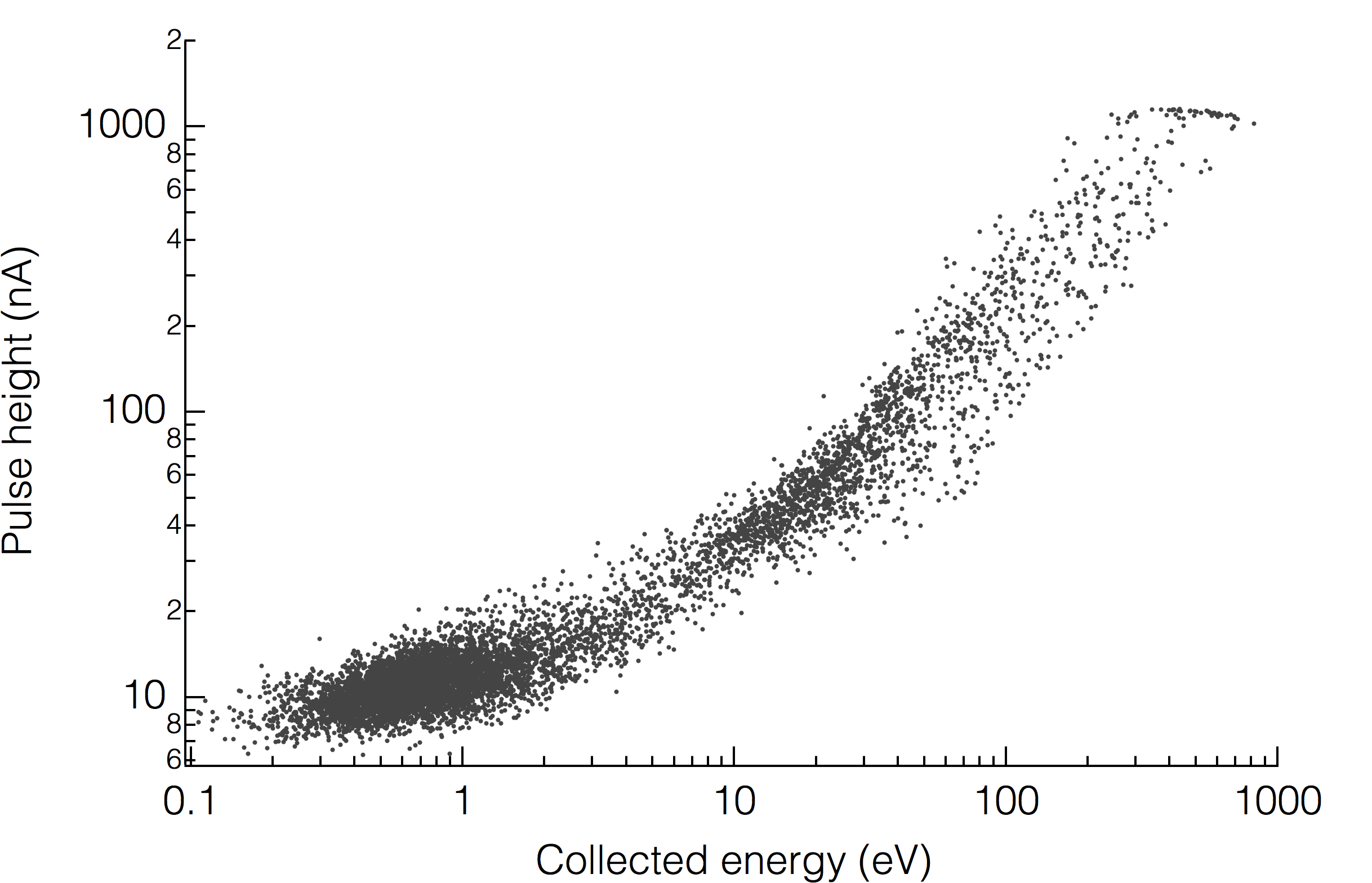}\label{fig:no-helium-scatter}}
	\\
	\subfloat[With helium]{\includegraphics[width=3.25in]{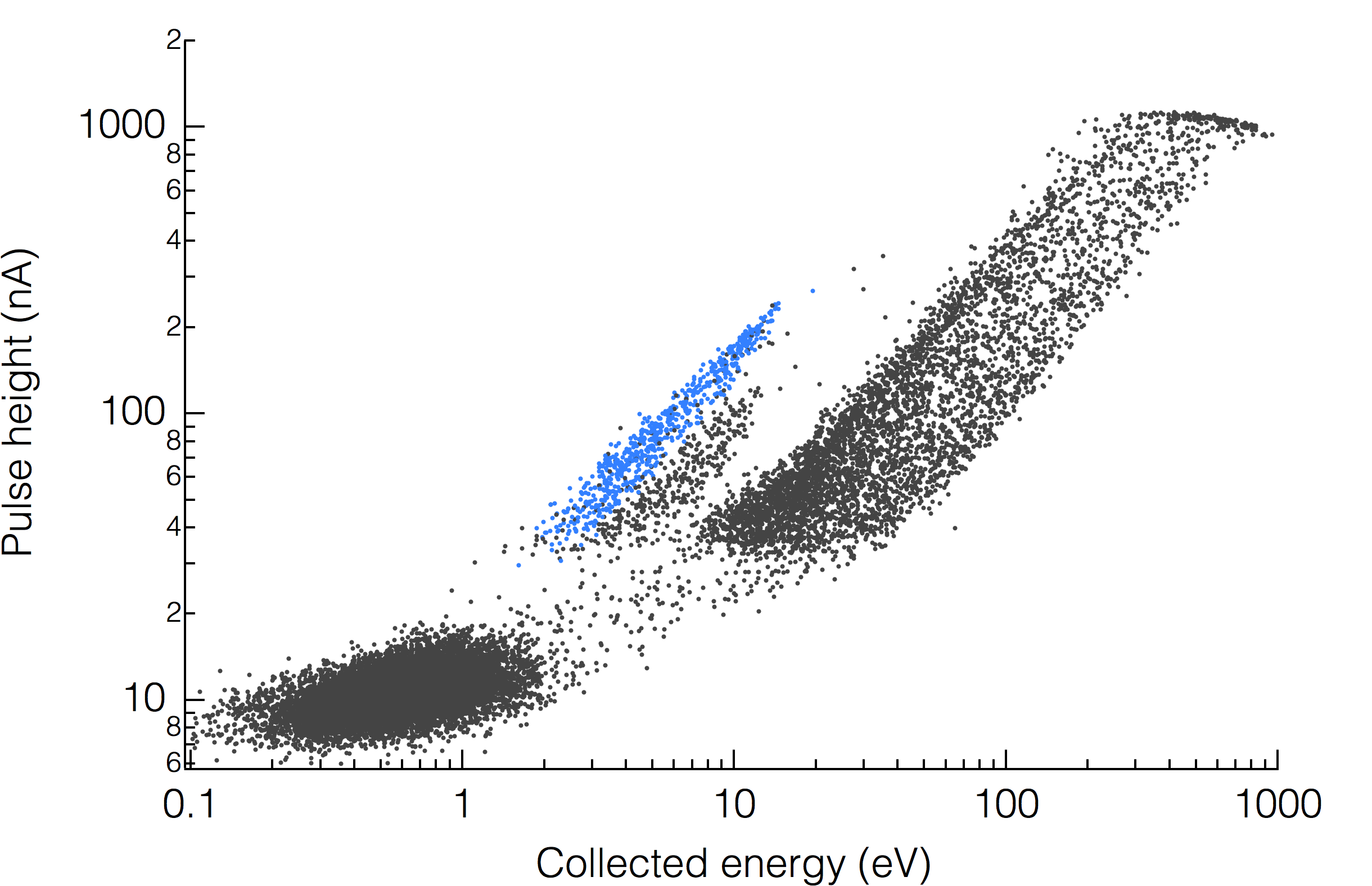}\label{fig:na22-scatter}}
	\caption[TES response to gamma-rays with and without superfluid helium]{(a) Pulse heights vs. collected energy recorded with the TES while an empty chamber was subject to a $^{22}$Na gamma-ray source. (b) The same, except in this case the chamber was filled with superfluid helium. Blue points correspond to direct excimer detection events in the TES (both singlet and triplet). The ``extra'' population located just below the blue points, but not present in (a), is due to energy deposited near the TES, but not directly in it (i.e. leads, substrate nearby, etc.).}
	\label{fig:gamma-scatters}
\end{figure}

When using the $^{22}$Na source some fraction of the gamma-rays will Compton scatter within the Si substrate rather than in the He. Such substrate energy depositions were observed, and excluded from the analysis using a pulse shape selection, which takes advantage of the relatively slow diffusion of phonons within the substrate, and their relatively weak coupling to the TES. Figure~\ref{fig:gamma-scatters} shows two scatter plots of pulse height vs. pulse area for events measured under $^{22}$Na irradiation. The events in Fig.~\ref{fig:no-helium-scatter} were recorded with the chamber empty of helium, and the events in Fig.~\ref{fig:na22-scatter} were recorded after condensing He. In both sets of data, a population corresponding to large pulse energies (10\textemdash1000~eV) and large pulse heights (30\textemdash1000~nA) was observed; this is the substrate absorption signal (the cutoff near 1000~nA is due to detector saturation). Random noise triggers are included as a blob in the lower left of each plot. In Fig.~\ref{fig:na22-scatter}, two populations of much faster (larger height-to-energy ratio), lower-energy pulses may be observed. The upper population (colored in blue) results from direct absorption of energy in the TES from either a singlet photon or a triplet quench. The lower population is due to events near the TES (i.e. the leads, the substrate below the TES, etc.) All of the following data analysis focuses solely on the `direct hits' (blue), which are selected by windowing on events with time-constants that match the intrinsic TES time constant.

\begin{figure}[h]
	\subfloat[]{\includegraphics[width=3.25in]{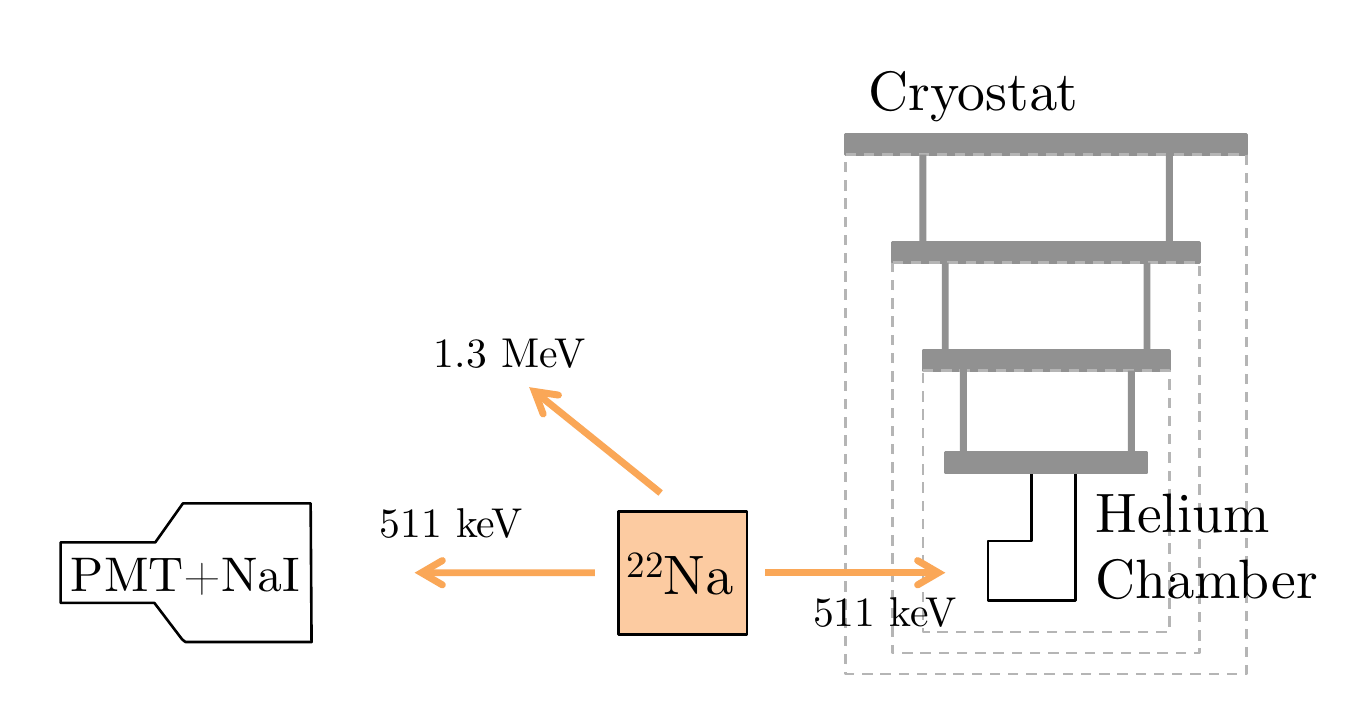}\label{fig:PMT-setup}}
	\\
	\subfloat[]{\includegraphics[width=3.25in]{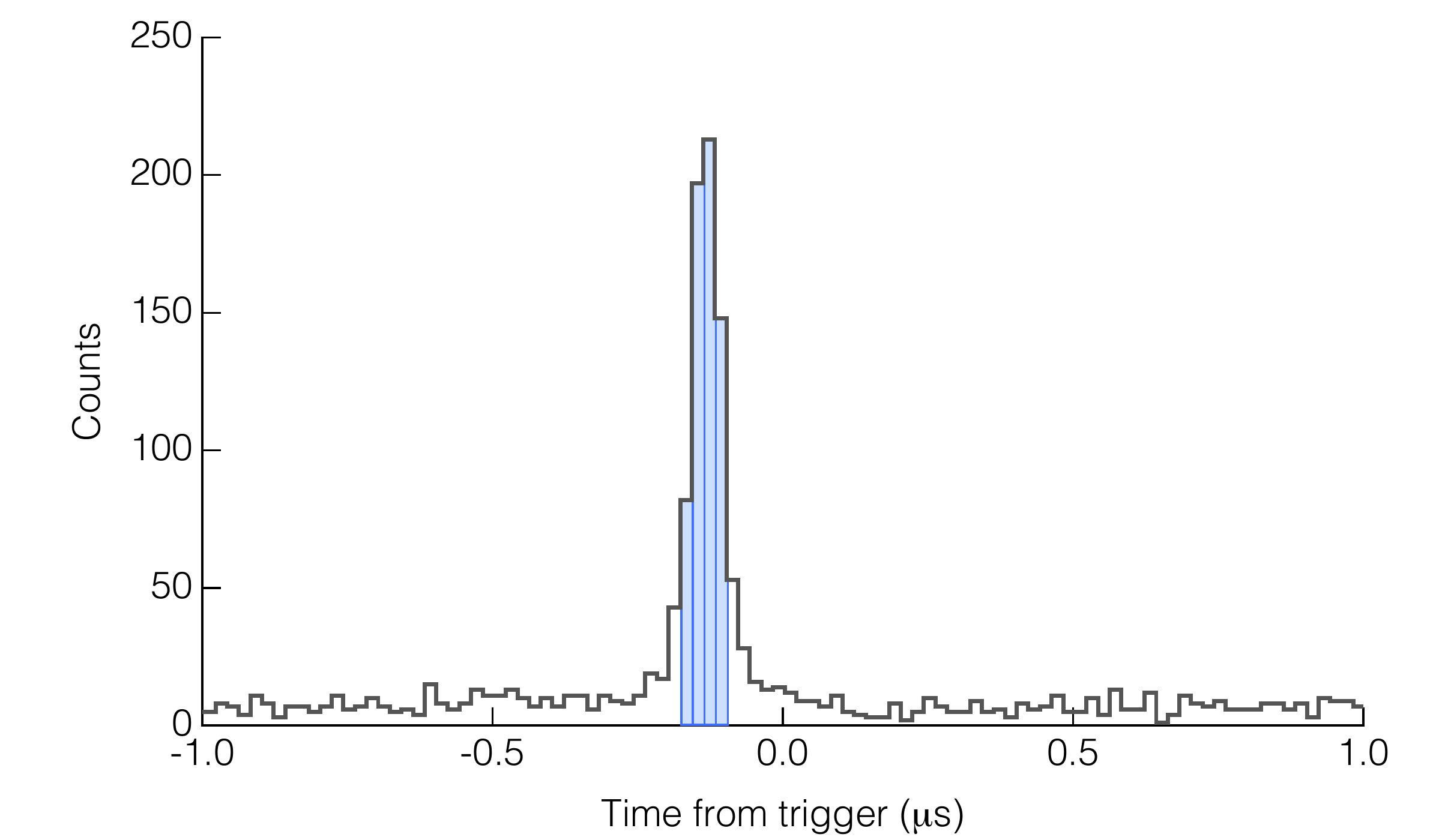}\label{fig:PMT-histogram}}
	\caption{(a) Experimental setup for coincidence measurements with a $^{22}$Na source. The PMT+NaI detector and the helium chamber are aligned relative to the radiation source such that the solid angles of illumination are matched. Any 511~keV gamma-ray that is incident on the chamber, will also produce a 511~keV gamma-ray that is incident on the PMT+NaI detector. (b) Histogram of the time difference between pulses detected in the PMT channel and pulses detected in the TES channel; substrate events have been removed. The shaded region is the cut used to make the red curve in Fig.~\ref{fig:na22-spectraTi}.}
\end{figure}

Each $^{22}$Na decay produces two counter-propagating 511~keV gamma-rays and a single 1.3~MeV gamma-ray. Thus, by adding a standard NaI scintillator coupled to a photo-multiplier tube (PMT) opposite the helium chamber, we may tag TES events coincident with a $^{22}$Na decay. Figure~\ref{fig:PMT-setup} illustrates the geometry for this setup. Singlet-decay photons are emitted promptly (ns scale) after the gamma-ray/electron recoil, whereas triplet excimers arrive at the TES delayed by a ballistic propagation time measured by \citet{Zmeev2012} to be $\sim$2~m/s. These dramatically differing timescales imply that any TES event which is coincident (within 250~ns) with a PMT signal must arise from the collection of a prompt photon, or a triplet state excimer that was created within a distance of less than 12~$\mu$m from the TES. The chamber is big enough that this small population of triplets will contribute fewer than one out of every fifty detection events and may thus be neglected.

The PMT output is continuously monitored by the same oscilloscope that monitors the TES. Whenever a TES signal triggers the oscilloscope, it collects 5~$\mu$s of data before the trigger and 45~$\mu$s after the trigger. The rising edge of any pulses observed in the PMT channel are recorded as delay times relative to the trigger time at $t=0$. Figure~\ref{fig:PMT-histogram} shows a histogram of delay times between TES events (at $t=0$) and PMT events. The large peak near zero delay is due to photon absorption events in the TES that were coincident with a gamma-ray detection in the PMT. The $\approx$100~ns offset from zero reflects the combined effects of cable-length delay and the placement of the TES trigger partway up the rising edge of the TES pulse.

\section{Results}
\begin{figure}
	\includegraphics[width=3.3in]{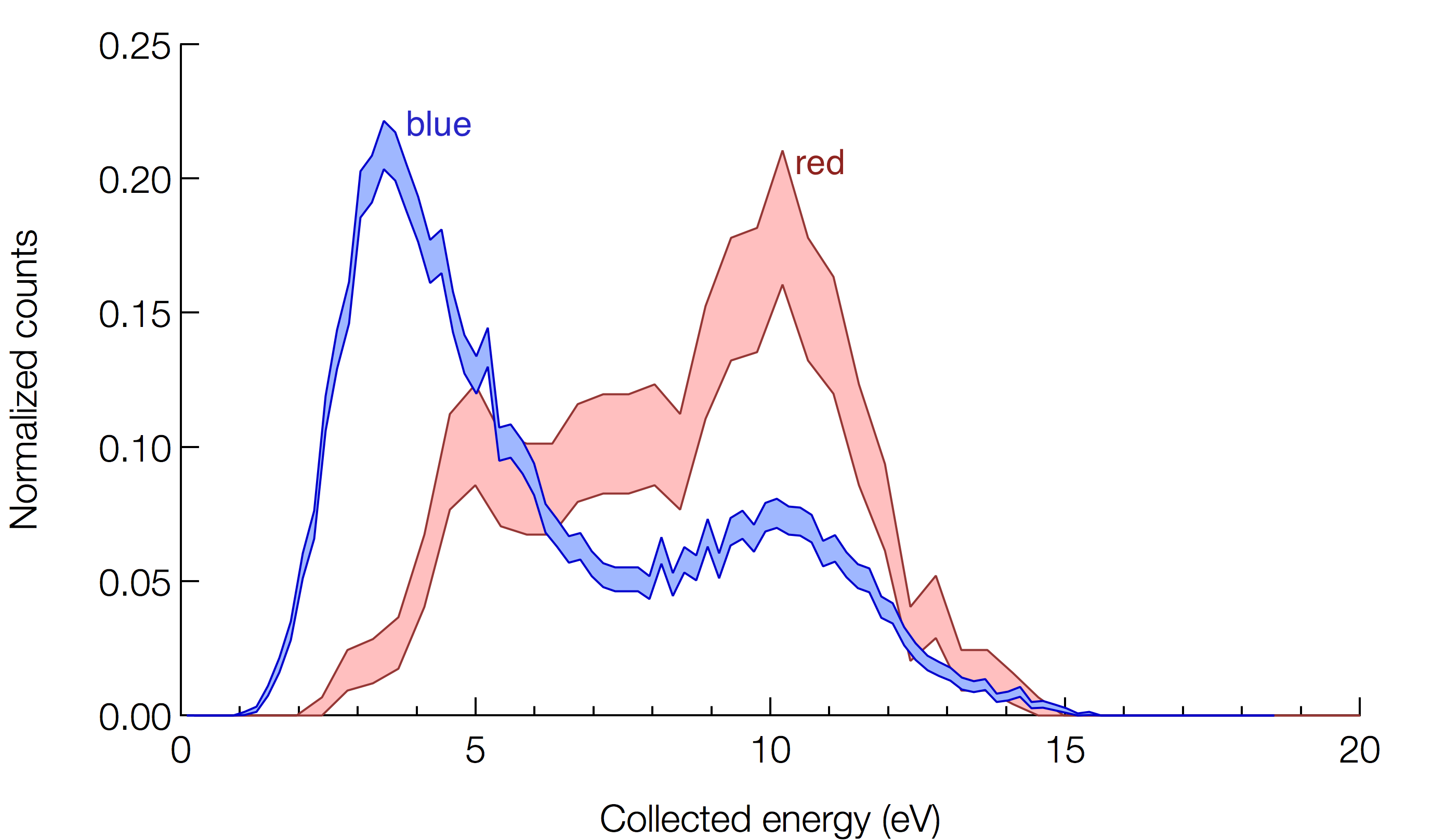}
	\caption{Red curve: detected emission spectrum of singlet He$_2^*$ decays (683 counts). Blue curve: total spectrum of all detected events (13\thinspace256 counts). Shaded area between lines indicates error, calculated as $\pm$ the square-root of the counts in each bin.}
	\label{fig:na22-spectraTi}
\end{figure}

Figure~\ref{fig:na22-spectraTi} shows two energy spectra created by binning $\sim$13\thinspace000 single excimer detection events by the energy measured with the TES. The blue curve is a spectrum of all the TES events due to irradiation by the $^{22}$Na source (the population of events colored blue in Fig.~\ref{fig:na22-scatter}). The red curve shows a spectrum consisting of only the coincident events (the events in the shaded region of the histogram in Fig.~\ref{fig:PMT-histogram}), which must arise from detecting singlet photons (plus a 3\% contribution from non-coincident events due to random coincidence). The shaded area indicates the error, calculated as the square-root of the number of counts in each bin. The two curves have been scaled such that their areas are each equal to unity. We hypothesize that the large peak in the blue curve near 3.5~eV arises mostly from the detection of triplet state excimer quenches on the TES surface, although it is clear that the photon-only signal (red curve) also has a small contribution in that energy range.

\subsection{Discussion}
\subsubsection{Coincident spectrum (red curve)}
The primary feature of the helium scintillation spectrum as measured previously by \citet{Stockton1970} is a peak centered at 15.5~eV with a full-width-half-maximum of about 3~eV. One would expect the red spectrum in Fig.~\ref{fig:na22-spectraTi} to look qualitatively similar, as it is due almost entirely to scintillation photons. Instead, the coincident spectrum peaks somewhat lower, near 10~eV, and shows a tail towards lower energies. It is possible that the detector response may not be linear from 2.65~eV (the calibration energy using blue photons) to this higher 15.5~eV range. We investigated TES linearity at low-energy experimentally by comparing our 2.65~eV calibration with a separate 0.8~eV calibration; results were consistent with linearity in this range (an identical fraction of photon energy appeared in the TES electron system during the absorption of a single photon: 67~\%). Calibration at higher photon energies was not performed as we did not have access to such a photon source. 

In order to estimate the efficiency of our TES at higher energies we calculated the energy lost during a photon absorption according to \citet{Kozorezov2013}. When the detector absorbs a single photon all of the energy is initially contained in a single excited photo-electron. This excited electron transfers its energy to the electron system, which arrives at a thermal distribution and then cools off through phonon-scattering. This is a four-step process. Step 1: The initial photo-electron shares its energy with other electrons within a radius of about 20~pm via electron-electron interactions until the mean electron energy in the hot spot is about 800~meV, independent of initial energy (this is a material property of Ti). This process takes of order 100~fs. Step 2: The athermal electrons then shed phonons until the rate of phonon emission roughly matches the rate of phonon absorption, which happens at a mean energy per electron of about 4~meV for titanium and takes a time of order a few picoseconds. Step 3: The hot spot of electrons shares its energy with the rest of the system through electronic interactions and arrives at a thermal distribution within a few nanoseconds. Step 4: The thermal electrons cool through phonon scattering over a time-scale of order 1~$\mu$s for our device. This characteristic time is shortened to about 700~ns in our experiment through electro-thermal feedback due to the voltage bias. During Step 2 of this process a large amount of the initial photon energy is converted into athermal phonons. For a thin film device (thinner than the athermal phonon mean-free-path) the chance of energy loss due to phonons escaping into the substrate is high. We have calculated that for our device this loss should be about 30~\% of the initial energy, a value that is consistent with our experimentally determined device energy efficiency of 67~\%. However, we find that this efficiency does not depend on initial photon energy (unless the initial photon energy is smaller than 800~meV in which case the energy loss will be less), and so can not explain the apparent non-linear effect we observe at 15~eV.

There are at least three remaining possible additional channels for energy loss during a photon event. At the beginning of Step 4, the hot electrons have a mean energy of about 4 meV, which is roughly 20 times higher than the aluminum superconducting energy gap ($\Delta$). The time for an electron to diffuse from the center of the TES to the edge ($\tau = L^2/D$, where $L$ is the distance to the contact and $D$ is the diffusion coefficient) is about 80~ns, which is of the same order as the time it takes the hot spot to settle into a thermal distribution below the Al energy gap. This allows for the possibility that some hot electrons could escape out the Al leads prior to thermalization. The energy lost in this process should be a linear function of the initial number of hot electrons present, and so a linear function of photon energy. Our two-point calibration (0.8~eV and 2.6~eV) does not exhibit any significant efficiency variation and so we discount this loss mechanism.

The second energy loss channel is the photo-electric effect. At the beginning of Step 1, a photo-electron with energy greater than the Ti work function (4~eV) may simply escape the TES entirely. This can happen either immediately or after sharing some of its energy with the electron system. \citet{Mancini1981} show that for thin films, the escape probability is proportional to $1/d$ where $d$ is the film thickness. Studies of photo-electron yield for micron-thick Ti films give yields as large as 20~\% \citep{Cairns1966}, suggesting that a 100~\% yield for a very thin film is not unreasonable. When a photo-electron escapes, it leaves a hole below the Fermi surface that is filled by an electron from the Fermi level, releasing the remainder of the energy into the TES according to the four step process above. \citet{Walker1962} studied the energy spectrum of photo-electrons emitted from a gold film during illumination from a 14~eV source and found a large fraction of escaping photo-electrons with very low energies ($\sim$3~eV) suggesting that they came from deep in the Fermi sea. If a similar process were occurring in our detector with 100~\% conversion rate, it would explain the apparent drop in efficiency at 15~eV that we observe.

The third loss channel relies on the TES having surface oxidation. A 15~eV photon is large enough to remove an oxygen ion through a photon-assisted Auger process~\citep{Hanson1981}. The characteristic energy of the ejected O$^+$ is about 4~eV, so for a 15~eV photon, 11~eV would be deposited in the TES. We expect the occurrence of this phenomenon to be quite low, but include it here for completeness.

Finally, we note that \citet{Cabrera1998} characterized a thin-film tungsten TES at energies from near-IR up to 3.5~eV and found that while their device was linear up to 3~eV, it experienced an onset of decreased efficiency above 3~eV (see Fig.~4, inset, in their paper). This motivates further study of the energy loss of thin-film TES devices in the vacuum-ultraviolet energy range where photo-electric and Auger processes may be important.

\subsubsection{Total spectrum (blue curve)}

The total spectrum (collected without regard to coincidence in the PMT channel) also shows the 10~eV peak that we attribute to absorption of singlet-state excimer decay photons; the coincidence trigger efficiency was low so we could not construct a data selection free from this prompt contribution. A second peak is observed near 3.5~eV, and we conclude that this peak represents the direct detection of triplet-state excimers after a propagation delay.

\begin{figure*}
	\subfloat[Just before coupling]{\includegraphics[width=2.75in]{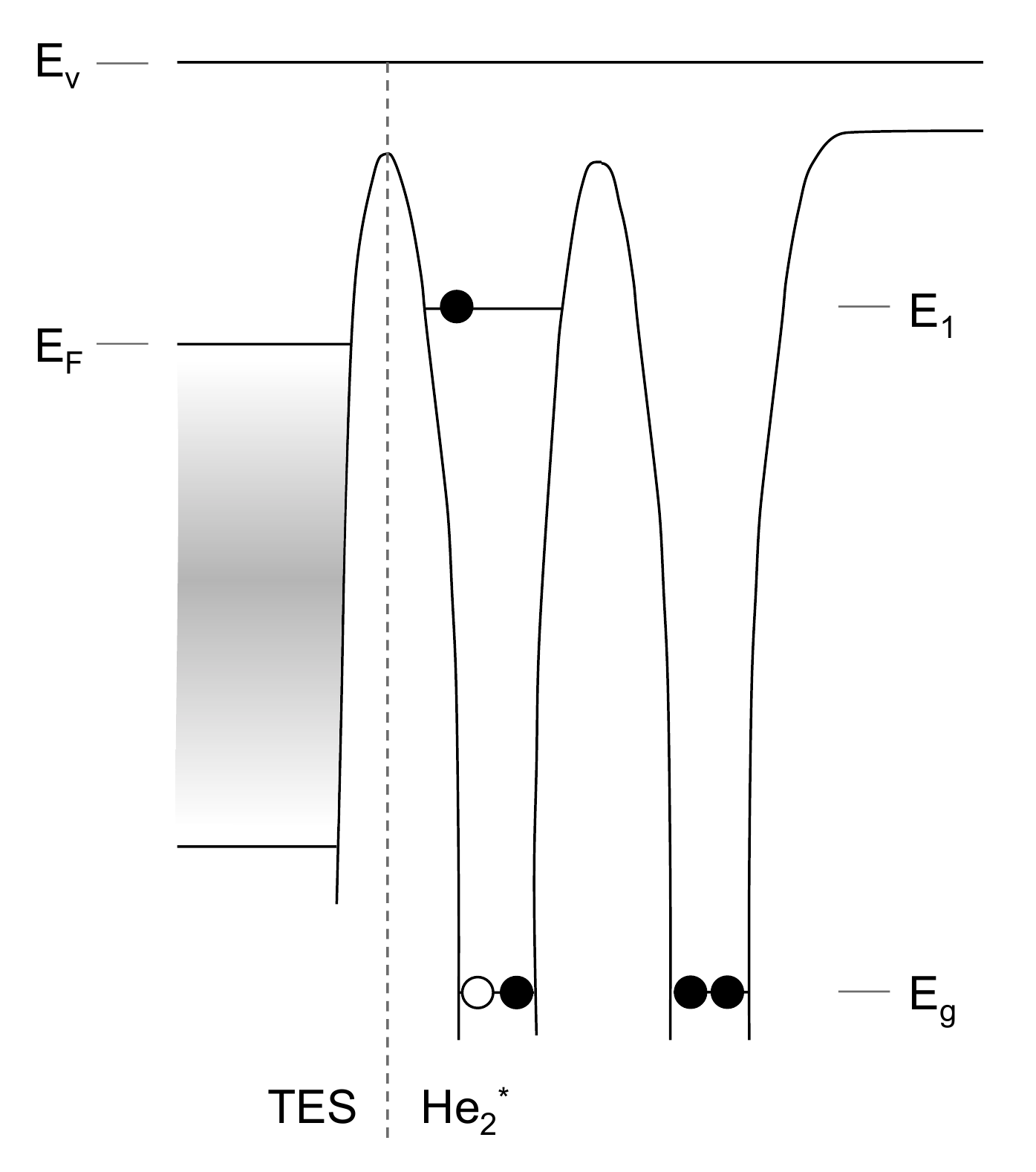}\label{fig:excimer-coupling1}}
	\hfil
	\subfloat[Just after coupling]{\includegraphics[width=2.75in]{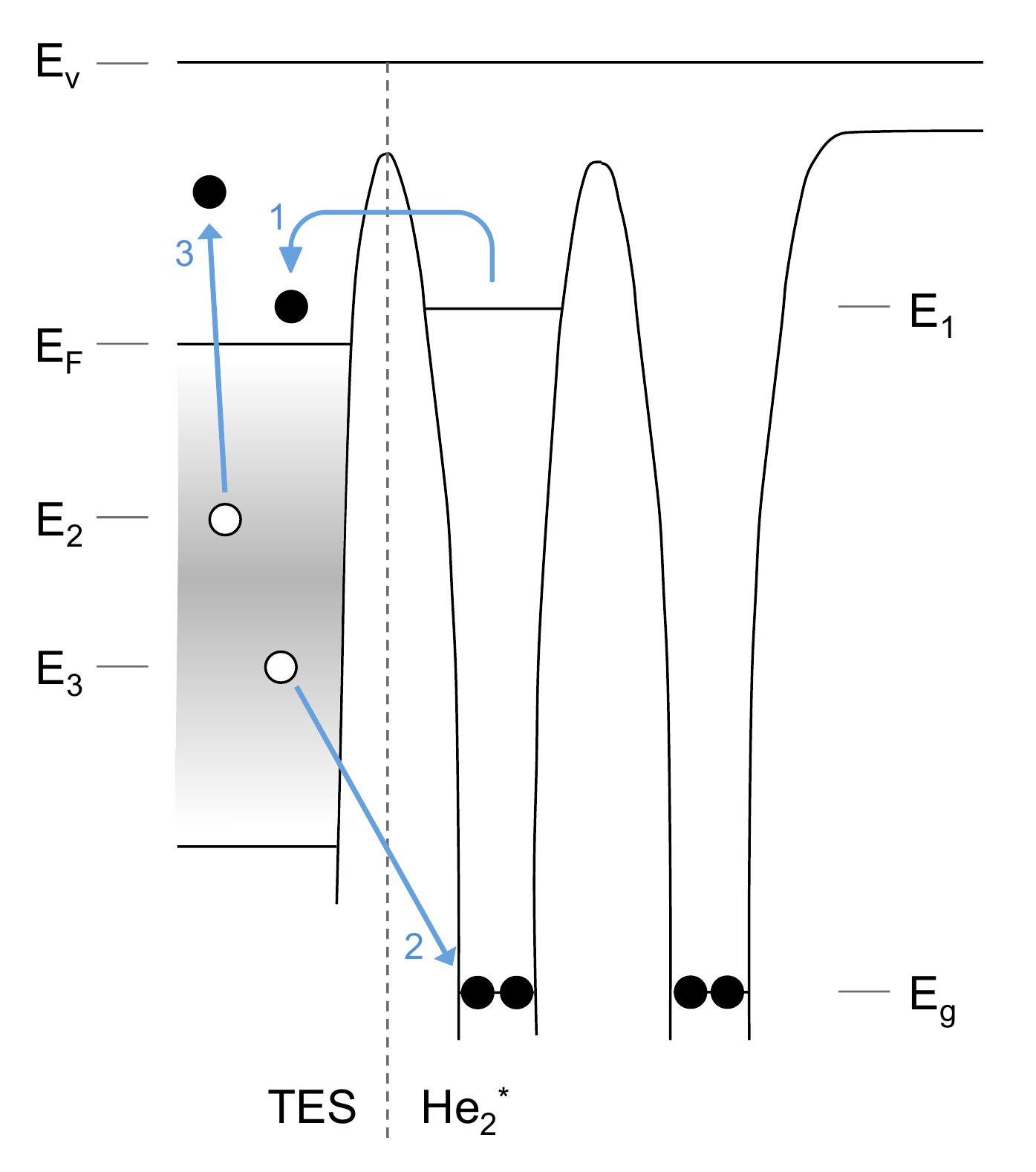}\label{fig:excimer-coupling2}}
	\caption[Process by which a triplet excimer couples to a TES]{(a) The triplet state excimer is to the right of the dashed line (one He atom in the ground state, and one in an excited state) and the TES is to the left of the dashed line. The horizontal axis is distance, and the vertical axis is energy. $E_\mathrm{v}$ is the vacuum energy and $E_\mathrm{F}$ is the Fermi energy of the TES. $E_\mathrm{g}$ is the ground-state energy of the excimer. The gradient region depicts an arbitrary non-uniform density of states below the Fermi level in the TES. $E_1-E_\mathrm{g}$ is the excimer energy and is about 15~eV. $E_1$ is roughly 4~eV below $E_\mathrm{v}$, but shifts upwards as the excimer approaches the TES surface. (b) The processes involved in the excimer quenching on the detector (blue arrows) is numbered in order of occurrence 1: the excited electron tunnels into a free state in the TES, and relaxes to the Fermi surface. 2: An electron from the TES Fermi sea fills the empty ground state of the helium molecule, and the molecule splits into two atoms. 3: An Auger electron is promoted from within the TES Fermi sea, and has energy $E_3-E_\mathrm{g}$. Finally, electrons relax from the Fermi surface to fill the two empty states in the TES (this is not pictured). Energies $E_2$ and $E_3$ are arbitrary as these two electrons may source from anywhere in the TES Fermi sea.}
	\label{fig:excimer-coupling}
\end{figure*}

Deexcitation of single He$^*$ atoms through interaction with solid surfaces has been studied for the purpose of probing the electron density of states in a material's surface (metastable deexcitation spectroscopy). See, for example, \citet{Harada1997}, \citet{Bonini}, or \citet{Trioni2005}. He$_2^*$ deexcitation (or quenching) at a surface is expected to proceed in a similar fashion. Figure~\ref{fig:excimer-coupling} depicts the manner in which an excimer quenches at the surface through electron exchange, coupling a fraction of its 15.5~eV excitation energy into the TES surface. The process is as follows:

\begin{enumerate}
	\item As the excimer comes within angstrom-scale distance of the TES surface, the excited electron resonantly tunnels into an empty state (at energy $E_1$) above the metal's Fermi level ($E_\mathrm{F}$), leaving the excimer charged as He$_2^{*+}$. This electron at $E_1$ then scatters down to the Fermi energy, releasing an energy ($E_1-E_\mathrm{F}$) into the TES. For this `resonant ionization' process to occur, the work function of the TES surface must be greater than $E_\mathrm{v}-E_1$ where $E_\mathrm{v}$ is the vacuum energy. It should be noted that both the surface and He$_2^*$ electron energies vary as a function of surface-He$_2^*$ separation (excimer electron energies increase slightly as distance decreases).
	
	\item An electron from the metal surface then tunnels in the opposite direction, filling the vacancy in the ground state to neutralize the He$_2^{*+}$.  The dimer separates.
	
	\item An electron from within the Fermi sea gains the energy released in the previous step ($E_3-E_\mathrm{g}$) through an Auger process. This may be enough energy to cause the electron to escape the TES. This and the previous step are often termed `Auger neutralization'.
	
	\item Finally, electrons from the Fermi surface cascade down to fill the two holes left by the neutralizing electron and the Auger electron. These two energies are $E_\mathrm{F}-E_3$ and $E_\mathrm{F}-E_2$. If the Auger electron from the previous step remains within the TES, it will deposit the rest of its energy, $(E_3-E_\mathrm{g})-(E_\mathrm{F}-E_2)$, in the TES as it relaxes to the Fermi level.
\end{enumerate}

In this process, $E_1-E_\mathrm{F}$ is always deposited in the TES. For a Ti TES, this energy is very nearly zero. The two electrons from steps 2 and 3 can come from any filled state in the TES, and so the density of states plays a large role in determining $E_2$ and $E_3$. There is also no reason why $E_2$ should be larger than $E_3$. This process effectively maps out the density of states near the Fermi level, and is nearly the inverse of metastable deexcitation spectroscopy. In that process the Auger electron is collected when it escapes from the surface, whereas in this case, the TES collects the energy that does not escape. Thus, the spectral shape will be heavily influenced by both the density of states near the Fermi surface of the TES and the detector's response to input energy.

This explanation relies on the efficient escape of Auger electrons from their origin in the outermost atomic layer of the TES. While there are no measurements of this escape probability specifically for He$_2^*$ metastable deexcitation on Ti, there are a variety of measurements on other metals for the similar He$^*$ deexcitation process. The Auger electron escape probability has been observed to range from $\approx$0.45 to $>$0.90 depending on metal and surface treatment (generally higher for air-contaminated surfaces), consistent with our observations \citep{Hotop1996}.

\section{Conclusion}
We have demonstrated the calorimetric detection of individual He$_2^*$ excimers in two distinct channels with a TES immersed directly in superfluid helium. The short-lived singlet states are visible through their $\approx$15~eV scintillation photons, well above threshold of the device. The long-lived triplet states are also detectable through their few-eV energy deposition upon arrival at the TES surface. The two signals are distinguishable (in aggregate) by their spectral shapes.

The energy deposited during the interaction of the triplet excimer with the TES surface is expected to be highly dependent on the TES surface's electronic density of states. In our case, we posit that the oxidized Ti surface allowed the efficient transfer of a fraction of the excimer energy into the surface by taking advantage of bands located between 2~eV and 5~eV below the Fermi energy. In a future version of this experiment, this density of states could be engineered to generate a signal of higher efficiency or fidelity. The eventual application of these new excimer detection principles towards instrumenting a large volume with high excimer detection efficiency will depend on applying them to larger-area calorimeters (see \citet{CRESST2005} or \citet{Pyle2015}).

One promising application for our detector concept lies in studies of quantum turbulence. Given the large (96~nm) trapping diameter for He$_2^*$ molecules on quantum turbulence vortices \citep{Zmeev2013}, high-efficiency detection of these excimers should enable sensitive monitoring of quantum turbulence density by measuring the He$_2^*$ flux passing through a superfluid helium-filled volume. This technique would also allow for independent monitoring of the He$_2^*$ production from sharp tungsten tips held at a high voltage and used to decorate quantum turbulence with He$_2^*$ excimers.

Finally, we note that there is a considerable uncertainty as to the exact physics underlying the spectral shapes presented in Fig.~\ref{fig:na22-spectraTi}. We have put forward a hypothesis that appeals to the surface electron density of states, the photo-electric effect, and Auger processes, but there is no detailed model. A full understanding is not possible within the limits of the current data, and a new set of experiments and devices will be required to obtain a detailed theoretical explanation of these promising first results.

\begin{acknowledgments}
We would like to thank Prof.~M.~Devoret for the loan of a dilution refrigerator; J.~Cushman for drafting expertise; C.~Matulis for circuit-board design; Dr.~L.~Frunzio for fabrication advice; Prof.~R.~Schoelkopf and Prof.~M.~Hatridge for cryogenics expertise; Dr.~C.~McKitterick, Dr.~Z.~Leghtas, and S.~Touzard for helpful discussions; and the Gibbs Machine Shop for making experimental hardware. Facilities use was supported by YINQE and NSF MRSEC DMR-1119826. We also acknowledge support from the National Science Foundation under NSF DMR-1007974.
\end{acknowledgments}

\appendix*

\section{Fabrication}
\label{apx:fabrication}
The TES was fabricated on a bare (unoxidized) high-resistivity silicon wafer. All the metal depositions were accomplished using electron beam (E-beam) evaporation in a commercial evaporator from Plasma Systems. The patterning was done using electron-beam lithography in a Raith EBPG 5000+ system.

\begin{enumerate}

\item{\textbf{Alignment marks:}
The wafer is cleaned, and a bilayer of polymethylmethacrylate (PMMA) E-beam resist is applied. A pattern of $20\times20$~$\mu$m$^2$ squares are written via E-beam, and then 400~nm of Cu is evaporated and lifted-off to define the alignment marks. These marks are used to precisely position the etch window over the region destined to become the TES.}

\item{\textbf{Ti/Al bilayer:}
A bilayer of PMMA is applied. The outline of the detectors, wiring, and absorbers is written via E-beam. Then a bilayer of 15~nm of Ti and 300~nm of Al is evaporated and lifted-off.}

\item{\textbf{Al etch:}
A single layer of PMMA is applied. Windows are opened in the resist via E-beam lithography directly over the areas that are to become the TES. The exposed Al is then etched down to the Ti, defining the detectors (Ti is a natural etch stop for the etchant used).}

\item{\textbf{Aperture:}
The device is coated with a 1~$\mu$m thick layer of PMMA. Then a bilayer of 50~nm of Cu and 50~nm of Al is evaporated. An additional single layer of PMMA is applied, and a window is opened up over the TES. The Al and Cu are both etched away, leaving the 1~$\mu$m layer of PMMA. This layer is then developed away, as it received a weak dose of E-beam current during the earlier E-beam write.}
\end{enumerate}

\bibliography{HePaper}

\end{document}